\documentclass[10pt,twocolumn]{article}

\usepackage{graphicx}
\usepackage{subcaption}
\usepackage{amsmath,amssymb,amsfonts}
\usepackage[bookmarks=false]{hyperref}
\usepackage{url}
\usepackage[gen]{eurosym}
\usepackage{authblk}

\begin{document}
\title{Novel paradigms for advanced distribution grid energy management}
\author[*]{Jos\'e Horta}
\author[*]{Daniel Kofman}
\author[**]{David Menga}
\affil[*]{Telecom Paristech, 23 Avenue d'Italie, Paris, France\\ Emails: \{jose.horta, daniel.kofman\}@telecom-paristech.fr}
\affil[**]{EDF R\&D, EDF Lab Paris-Saclay, 91120 Palaiseau, France\\ Email: david.menga@edf.fr}
\date{}

\maketitle

\begin{abstract}
The electricity distribution grid was not designed to cope with load dynamics imposed by high penetration of electric vehicles, neither to deal with the increasing deployment of distributed Renewable Energy Sources. Distribution System Operators (DSO) will increasingly rely on flexible Distributed Energy Resources (flexible loads, controllable generation and storage) to keep the grid stable and to ensure quality of supply. In order to properly integrate demand-side flexibility, DSOs need new energy management architectures, capable of fostering collaboration with wholesale market actors and prosumers. We propose the creation of Virtual Distribution Grids (VDG) over a common physical infrastructure, to cope with heterogeneity of resources and actors, and with the increasing complexity of distribution grid management and related resources allocation problems. Focusing on residential VDG, we propose an agent-based hierarchical architecture for providing Demand-Side Management services through a market-based approach, where households transact their surplus/lack of energy and their flexibility with neighbours, aggregators, utilities and DSOs. For implementing the overall solution, we consider fine-grained control of smart homes based on Internet of Things technology. Homes seamlessly transact self-enforcing smart contracts over a blockchain-based generic platform. Finally, we extend the architecture to solve existing problems on smart home control, beyond energy management.

\end{abstract}

\section{Introduction}
The increasing deployment of distributed renewable energy sources (RES), like residential solar panels attached to the distribution grid\footnote{The electricity grid is structured hierarchically into transmission grid level carrying electricity at high voltages across long distances and distribution grids that deliver medium/low voltage electricity in shorter distances.}, and the growth in adoption of electric vehicles (EV) are posing severe challenges to Distribution System Operators (DSO), e.g.: in terms of quality of supply, congestion, voltage variations and on the protection system. In order to cope with these challenges DSOs will increasingly rely on flexible\footnote{Flexibility, in this context, can be defined as the capability to adapt demand and/or injection flows of electricity to the grid by adapting consumption patterns, controlling generation output or storing energy.} Distributed Energy Resources (DER), such as flexible loads, controllable generation, or storage resources, to keep the grid stable and to ensure quality of supply. Nevertheless, existing distribution grid energy mangament mechanisms are not adapted for such evolutions. In order to properly integrate demand side flexibility distribution grids need new energy management architectures, capable of fostering coordination and collaboration among wholesale market actors, DSOs and prosumers\footnote{Prosumers represent the evolution of passive consumer role into a pro-active participation on grid activities.}.

We consider an end to end architecture that, whenever necessary and convenient for players, can integrate the control of edge devices including customer appliances. The Internet of Things (IoT) paradigm promises to enable connected devices to identify themselves, describe their capabilities, discover each other, and self-organize in order to provide innovative services. In particular, in the context of Smart homes and Smart grids, it will enable customers to seamlessly play a more active role, facilitating providing flexibility to the grid.

Such flexibility is already being exploited at wholesale electricity markets through recently adopted market mechanisms and new roles. Flexibility is aggregated into Virtual Power Plants (VPPs) by Aggregators and offered in substitution of expensive and polluting power plants at the energy markets. Balancing Responsable Entities (BRE)(such as Utilities) can also use demand side flexibility and transact blocks of energy among them to optimize their participation on the market. Transmission System Operators (TSO) also leverage flexibility for balancing the grid in real time through the Balancing Mechanism (or tertiary reserve).

While wholesale markets and TSOs have adapted to leverage demand side flexibility, DSOs continue reinforcing and extending the infrastructure as the main lever to cope with distribution grid challenges, which is currently hindering RES deployment pace.

The use of flexible DER will enable to reduce/defer infrastructure investments and to gain fine grained control over infrastructure and services. But, at present, storage resources and Building Energy Management Systems follow a stand-lone approach, focusing on auto-consumption and local energy efficiency, and therefore may worsen the negative impacts on the distribution grid because of the lack of coordination among actors.

The objective of this work is twofold. First, to provide the architecture building blocks for new distribution grid Energy Management Systems (EMS), capable of fostering such coordination among actors. The architecture should allow DSOs to better control the power balance and quality of supply of the distribution grid. In particular, locally balancing distributed RES production with demand flexibility has the potential to alleviate congestion, reduce losses and consequently augment variable RES hosting capacity without further infrastructure reinforcements/extensions. Second, providing means to implement the novel building blocks proposed in this architecture. In particular, such implementation requires to solve issues at the household level of the architecture, related to conflicting transactions that may hinder the control of flexible DER. Such conflicts arrive when controllers from different domains (energy, health, security, comfort) share the same environmental scope or control simultaneously the same edge devices.

The paper first addresses the main design objectives and requirements of new generation distribution grids' EMS and analyzes current approaches for coordinating the allocation of flexible DER: Microgrids and VPPs. Then we propose the concept of Virtualized Distribution Grid (VDG) as a new paradigm facilitating distribution grid management. We focus on a residential VDG. We propose a hierarchical agent-based architecture capable of providing Demand Side Management (DSM) services through market-based resource allocation mechanisms. We envision the distributed implementation of several markets to enable households to transact their surplus or lack of energy and the demand flexibility budget with neighbours, utilities, aggregators and DSOs. The infrastructure that enables market exchanges is based on blockchain technology and self-enforcing smart contracts\footnote{http://szabo.best.vwh.net/}. Finally, we address the problem of conflicts among controller agents at the household level. The solution relies on an extension of the blockchain-based platform, which enables the implementation of Event-Condition-Action (ECA) control schemes on-premises\cite{Wong_anactive}, improving security and privacy compared with cloud-based ECA implementations.

\section{Distribution grid Energy Management System}
\subsection{Objectives of a distribution grid EMS}
The main goal of a distribution grid EMS is to enforce efficiency, reliability and quality of energy supply, for which leveraging DER flexibility through DSM mechanisms will be a main technological enabler. Thus, a distribution grid EMS architecture needs to enable the following functionalities:

\paragraph{Local energy balancing} Leverage coordination of DER to balance local production at the distribution level (neighbourhood/village). This has the potential to alleviate congestion, reduce losses and consequently augment variable RES hosting capacity. 
\paragraph{DSO ``Reserves''} Enable DSO to influence flexible DER behaviour for collaborating with voltage control and congestion management.
\paragraph{Wholesale Market participation} Maintain collaboration with global objectives for grid balancing, through actors participating in the wholesale market, such as suppliers and aggregators.
\paragraph{Value to customers} The prosumer side of the participation on all the previous mechanisms needs to be taken into account and the allocation of DER to such mechanisms needs to be optimized for customer goals and policies; e.g.: minimizing impact on privacy, ease of life and comfort. 

\subsection{Requirements for a distribution grid EMS}
The requirements for a distribution grid energy management architecture are the following:
\begin{itemize}
	\item Reliability/Fault Tolerance - In order to leverage DER flexibility, any solution that tends to distribute the intelligence will need to maintain system reliability and fault tolerance. In this context, the EMS architecture must continue working under communication errors, or under presence of node failures or byzantine nodes.
	\item Flexible and Extensible - Flexible and extensible implies Plug and Play capabilities, which enable seamless integration over time \cite{uGridControl}, i.e.: handle continual arrival and departures of appliances, resources and control agents at different levels of the architecture.
	\item Scalability - There are a multitude of DERs, spawned over distant geographic areas, owned by different actors. Various market players, each one with its own objectives, are interested on leveraging those resources at different time scales. The EMS architecture needs to cope with the scalability issues in terms of time, scope and heterogeneity of resources and involved actors.
	\item Trustworthiness - The actors in charge of the EMS or the energy management mechanisms and their implementation need to be trusted: This implies Transparency, Security and Privacy on the transactions among participating actors. With respect to transparency, visibility over DER at the distribution level is becoming increasingly important to avoid system imbalances. 
	\item Low Implementation and Maintenance Cost - Business models are not yet clear, as pricing schemes and flexibility value characterization are still being studied, so the costs imposed by the energy management mechanism must be minimized.
	\item Re-usability of Infrastructure and platforms across distribution grids (Energy, Water, Gas, Heat). Re-usability of infrastructure and technologies should be aimed whenever possible, as a mater of costs but also due to interoperability, cross resource services (Combined Heat and Power management), critical mass of experts, etc.
	\item Independent evolution of control strategies. The architecture should enable the independent evolution of control strategies at different levels (Household, Distribution Grid, System Level), to cope with the fast progress of technologies and control mechanisms. 
	\item Efficiency and equity. In order to maximize social welfare, energy management mechanisms should maximize the value extracted from flexible resources and provide a fair distribution across participants, according to the service definitions. Furthermore, economic efficiency is required to incentive actors participation.
\end{itemize}

\subsection{Current approaches for DER management}
\begin{figure}[tb!]
	\centering
	\includegraphics[width=3.5in]{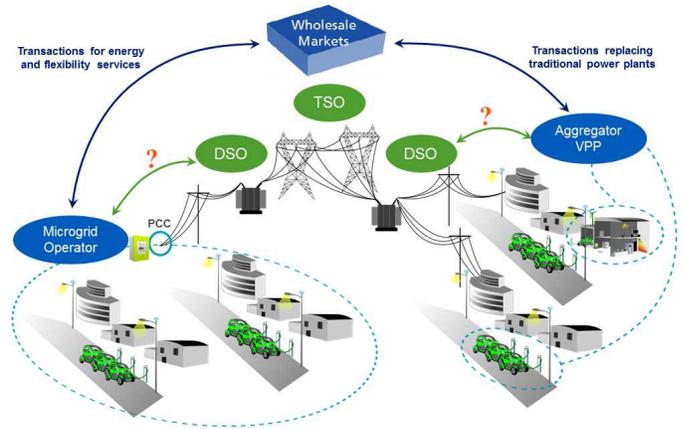}
	\caption{Microgrid and VPP approaches to DER management}
	\label{fig_ugridvpp}
\end{figure}

The two main approaches to manage flexible DER, both in literature and in practice, have been Microgrids and Virtual Power Plants (VPPs). Microgrids are defined as an entire partition of the physical infrastructure, with a Point of Common Coupling as unique interface with the hosting grid. This enables the possibility of isolation, but requires a unique actor to manage electricity flows exchanged with the hosting grid and economic transactions into the wholesale market (Figure \ref{fig_ugridvpp}). For this reason, Microgrids are a good model for a university campus or small communities. However, in the general case, there would be several actors (suppliers, aggregators, etc.) participating individually on wholesale markets by aggregating DER from the distribution grid. Thus, if we were to consider the Microgrid approach as the canonical method for distribution grid energy management, we would need to redefine the roles of DSO, aggregators and current BRE, and to adapt wholesale market rules accordingly.

On the other hand, VPPs are focused on wholesale market participation as a replacement of conventional power plants, and for this they aggregate DER flexibility from accross several distribution grids. VPPs are implemented through contracts with flexibility DER owners, obtaining the exclusive control of physical devices. In consequence, the corresponding DER are not available, neither visible to third parties. This lack of visibility may have negative impacts in grid balancing if there is no coordination among actors. VPPs play a major role in the participation of DER in wholesale markets but they do not have a clear role at the distribution level and may not take into account distribution grid constraints. Other papers have considered coordination among VPPs and DSO: \cite{caldon2004optimal} assumes that ``a sort of local market is established'' in order to be able to coordinate the concurrent operation of DSO and VPP through appropriate economic signals. While the Fenix project defines a Technical VPP which clusters RES from the same geographical area and would need tight coordination with the DSO \cite{TVPP}. To the extent of our knowledge, current literature about VPP has not addressed a general way of coordinating DER participation on wholesale markets with distribution grid needs.

\section{Introducing the Virtualized Distribution Grid concept}
Current approaches fail to take into account the coordination at the distribution grid level across multiple actors as they focus on the aggregation of resources for participating on the wholesale market. Thus, we propose a new architecture for distribution grid energy management, with the goal of coordinating the usage of flexible energy resources among different actors. The approach is compatible with microgrids and VPPs, as it does not targets wholesale market participation, but requires Microgrids and VPPs to negotiate the access to flexible DER through the corresponding resource allocation mechanism. Any other future system aimed to leverage flexible DER in wholesale markets would need to interact with such coordination mechanisms. As a consequence of aiming neutral resource allocation mechanisms, the distribution grid can be transparently managed, offering visibility for all relevant grid actors. 

\begin{figure}[tb!]
	\centering
	\includegraphics[width=3.19in]{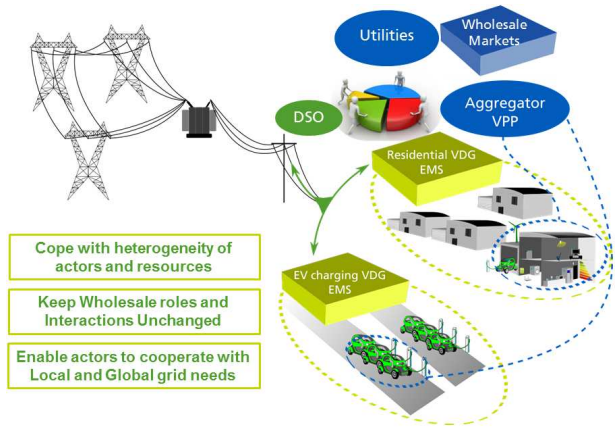}
	\caption{Virtual Distribution Grids approach to DER management}
	\label{fig_vdg}
\end{figure}

The resource allocation mechanisms need to cope with several types of flexible resources, with a wide scope of characteristics. Such resources are owned by various types of self-interested actors (residential, commercial and industrial prosumers, EV charging stations, energy service providers, aggregators, etc.), whose objectives are not necessarily aligned with distribution grid needs. To cope with the complexity of such mechanisms, we propose the creation of VDGs in charge of the energy management of subsets of resources by type and corresponding actors, over common distribution grid infrastructure. Each VDG will require different architectural building blocks and specific management policies and algorithms, e.g.: we could envisage a VDG for management of DER from residential prosumers in a neighbourhood and another for managing electric vehicle charging on a feeder (Figure \ref{fig_vdg}).

This is similar to the approach used on cloud environments to optimize the allocation of physical resources (processing power, storage or bandwidth), where a native hypervisor enables the creation of various virtual machines over a shared physical infrastructure. Every virtual machine needs an Operating System (OS) to access logical resources through the corresponding drivers. Additionally each OS will have algorithms to efficiently allocate available virtual resources to applications. An OS provides an execution environment to application programs, abstracting the low level complexity of resources from users and application developers (through user-spaces and Software Development Kits respectively).

Back to the energy management problem, the role of the DSO is analogous to the one of an hypervisor, in charge of keeping a coherent global behaviour between infrastructure management and energy management. DSOs would monitor physical resources and impose constraints to the each VDG\footnote{Note that there is no complete isolation among VDGs due to the correlation created by voltage variations. We are currently analysing mechanisms for coping with such issues.} resource allocation process or incentive behaviour of actors through a reserve mechanism. We can imagine that some VDGs would have more priority than others in the context of congested distribution lines, and by prioritizing, DSOs could even offer differentiated grid connection services.

The notion of OS on a VDG would be that of an EMS architecture\footnote{Note that the implementation of the EMS functions will run over an OS, but that is not what we are discussing here.} providing visibility and control over DER, by enabling its allocation to the different possible usages, in a similar way an OS allocates processing power to applications on a virtual machine. The main ``applications'' that will compete for DER should enable power balancing inside the VDG, participation in existing wholesale market mechanisms, and providing services to the DSO for voltage and congestion management. Actors foreseeing those uses will probably be competing for flexible resources in the same time-slots where the grid needs them more and flexibility has more commercial value. The EMS architecture needs to provide a cost-effective, transparent and secure coordination mechanism to allocate flexible resources to maximize social welfare; i.e.: to provide an equitable (e.g.: max-min fairness) and efficient (e.g.: Pareto efficient) allocation. In the following we will focus on the EMS of a particular type of VDG rather than on the role of DSO as an orchestrator of real and virtual resources. We will analyse general aspects of the design of an architecture capable of providing such an allocation of distributed resources.

\section{Design of a Residential VDG}
In this work we focus on the EMS of residential VDGs, i.e: on the management of flexible DER from residential prosumers in a neighbourhood/feeder. In particular, we are interested on locally balancing distributed renewable energy production by leveraging IoT technologies to communicate, monitor and control flexible residential DER. For the design of the residential VDG energy management architecture we will consider special attention to interoperability among architecture implementations, distribution of intelligence, security and privacy requirements, as these aspects play a major role on the viability of an IoT-based architecture.

\subsection{Hierarchical Architecture}
\label{subsec:controlschemes}
Centralized approaches for distribution energy management rely on a unique entity that must gather data, perform calculations and determine set points for every controlled resources. A centralized approach provides optimality at the expense of extensive communication and computing resources and lack of flexibility and adaptability. Furthermore, a central fully informed entity may not be available due to natural information asymmetries and selfish participants, i.e.: DER have different owners and decisions are made locally to optimize owner's objectives \cite{Lamparter}. Additionally, certain information is only available locally due to privacy concerns, such as presence information. Thus, it is hard for centralized control to handle customer policies based on context awareness, such as comfort or auto-consumption policies.

On the other hand, fully decentralized approaches that take decisions using local information without coordination can lead to synchronization and oscillation of the overall response. In particular, there exist already plenty of technologies enabling a more efficient usage of electricity and promoting auto-consumption for households that count on local PV production and/or storage facilities. These systems may not take the needs of the grid into account and affect the way utilities (and other BRE) balance customer demand and market supply, as their statistical models for load forecasting would not be valid any longer. Therefore, unless managed appropriately, such systems may create stronger and recurrent imbalances on the grid.

Even if the geographical extent of a residential VDG is not large, we still have a considerable number of controllable resources with stringent performance requirements, in terms of high-speed communication and computation of actions/set points for every unit\cite{uGridControl}. A hierarchical approach would drastically reduce such requirements for data gathering and processing, and would enable the independent evolution of control strategies and technologies by defining clear interfaces among levels (like through Application Programming Interfaces). Thus, we propose a mix of control strategies composing a hierarchy of three levels: the control inside each household, the allocation of prosumer resources at the distribution grid level and the Transport level that controls the overall grid balance.

We will focus on the two lower levels, as wholesale market and TSO control mechanisms, in spite of their continuous evolution, are more mature and reliable than the rest of control levels. Furthermore, we want to avoid requiring changes from the higher level other than small changes for the actors participating in the market, i.e.: on the way suppliers and aggregators gain access to flexible DER. 

\subsection{Household Control level}
Households represent the base of the hierarchical architecture for coordinating the allocation of DER resources to the different DR mechanisms depending on customer policies and economic incentives. In particular, security and privacy policies would lead to keep sensible data under the control of the customer to gain trust and acceptance. In order to enable such an empowerment, the distribution of data and the intelligence required for its treatment need to be designed accordingly. Considering a separate level for the energy management of a household enables such a distributed intelligence approach.

Each household will have an energy management function (HEMS) providing coordination among flexible loads, storage, local production and energy consumed from the global grid or from the VDG. Each HEMS will have its own knowledge and individual goals, and the capability to transact resources with the upper levels through a trading agent. The trading agent will abstract heterogeneity of household control implementations from the distribution grid level, i.e.: one agent representing the global goals of the smart home regardless of the coordination and control mechanism implemented inside each household.

\subsection{Distribution Grid Control level}
\begin{figure}[tb!]
	\centering
	\includegraphics[width=2in]{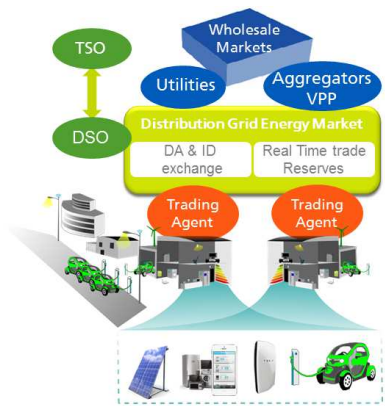}
	\caption{Hierarchical architecture for distribution grid energy management}
	\label{fig_hierarchitecture}
\end{figure}

In order to optimize the scheduling and dispatch of DER, coordination among households, DSO and other grid actors is mandatory. We envision the implementation of market-based mechanisms, enabling coordination and collaboration among trading agents representing households, DSO and other grid actors through the exchange of energy transactions (Figure \ref{fig_hierarchitecture}). Agents will transact energy resources with each other in order to meet household economic and comfort goals, while collaborating with local and global grid infrastructure operation issues\footnote{We assume the necessary regulation is in place to allow households to establish transactions among them for the exchange of energy.}. Markets have proven to be a suitable mechanism for resource allocation and control of autonomous selfish parties, and have already been tested for Distribution Grid Energy Management in the US, following the Transactive Energy approach \cite{transactive}, and in Europe under several demostrator projects generally involving microgrids \cite{alstom}\cite{retailmarkets}\cite{powermatcher}.

DSOs will implement agents representing the infrastructure needs from each VDG, for maintaining power quality and security constraints. This agent will be fed by the global distribution management system with day ahead operational requirements and by Supervisory Control And Data Acquisition (SCADA) system with information about grid conditions and reliability requirements for real time distribution automation.

BREs and other actors participating wholesale markets will be represented by agents on every VDG of the low voltage distribution level. This will enable those entities to gain feedback on the behaviour of the households comprised on their balancing domain and to better manage the aggregation of several sectors to provide services and to optimize their demand portfolio on the wholesale markets.

All previous processes need to take into account the economic scheduling and dispatch of DER resources in order to encourage participation on the EMS. Though business models are out of the scope of this work, we consider current economic signals are not favourable in most European countries for exploiting DER flexibility, but we assume they will evolve towards a more encouraging context.

\subsection{Wholesale Market-adapted Timescales}
Until now we have addressed the architecture hierarchy in terms of space or control areas and now we will consider the timescales in which energy resources would need to be allocated. We will consider three time scales, day ahead, intraday and real time, in order to adapt the allocation of flexible capabilities to the market structure and to ease the interaction among levels.

\subsubsection{Day-ahead process}
On the day ahead process, each HEMS will estimate the surplus or lack of energy for every hour of the next day, based on consumption and production forecasts. Then, the expected imbalance can be eliminated/minimized by scheduling controllable appliances or can be traded in the VDG market with other households or with actors participating on the wholesale market. The HEMS will decide the volume of electricity to be traded on the market with the corresponding price cap, by optimizing the scheduling of controllable appliances taking into account economic incentives (price of electricity supply, aggregator offers, feed-in tariffs, etc.) and customer policies (e.g.: comfort constraints). 

The trading agent representing each household will send the corresponding bids and asks to the VDG market, which will match  the orders determining the volume and price of equilibrium for each hour of the day. This process should finish earlier than the Day-ahead wholesale market, as the energy that was not balanced on the local market needs to be provided/absorbed through the wholesale markets or Over The Counter agreements, and the flexibility resources that were not traded on the local market remain available to the wholesale mechanisms. 

\subsubsection{Intraday process} 
The intraday process in each household will detect variations from the expected day-ahead load/production schedule and to allocate those gaps to the remaining local flexibility budget or to intraday VDG market mechanism. The VDG intraday market mechanism works in a similar way to the wholesale mechanism, where BREs can aggregate traded volumes and offer them to the balancing mechanism, the intraday energy market, or can exchange blocks with another balance responsible party in order to avoid paying penalties to the TSO. Thanks to this mechanism DSOs will have the capability to avoid congestion, reduce losses or control voltage by influencing demand-side energy behaviors, e.g.: by building a distribution grid reserve.

\subsubsection{Real-time process} 
In the real-time process, controllable loads need to fulfill the transactions they applied for in day-ahead and intraday, and the aggregators/BRE need eventually to execute balancing orders transmitted by the TSO in response to accepted bids on the balancing mechanism. In a similar way to previous processes, we will try to eliminate imbalances or to trade them into the market, but in this case the imbalance is measured with respect to an objective that depends on the mechanism to which resources were allocated. For example, if resources were allocated to an aggregator participating to the adjustment reserves, the objective will be to reduce/augment the energy consumed with respect to the last half an hour; while if resources were allocated to a distribution system reserve the objective may be to follow a curve provided by the DSO. The real time mechanism is not to be confused with \textit{Primary Control}, which is autonomous frequency and voltage control, and the fastest response to stabilize the grid under system dynamics. In this work we assume the physical resources attributed to the VDG are grid-connected and thus \textit{Primary Control} is assured at system level, i.e.: we do not consider isolation of physical resources.

\section{Control infrastructure and market implementation}
\subsection{Distribution grid control level}
We will describe the main ideas for the implementation of local balancing markets on the residential VDG, which represent the base for the other markets/auctions that aggregators, suppliers and DSOs can use to interact with flexible DER owners.

The implementation of such markets would need a third party on the role of auctioneer, for keeping an order book, matching orders at the end of the round/period, i.e.: defining the price and the volumes to be traded, and also for clearing and settlement, through which assets (in this case money and energy) are verified and the economic transactions are executed. The rounds/periods will depend on the market, as there will be several markets operating simultaneously; for instance we will have day-ahead, intra-day and real-time markets, where excess or lack of energy can be balanced locally.

For the local balancing market, we envisage the exchange of energy blocks among trading agents to be carried out in a call market, through a Multi-unit Double Auction (MDA) mechanism, where agents submit a bid or ask for orders, which are respectively the maximum, or minimum, price the agent is willing to pay to buy, or to accept to sell, a certain volume of energy on certain time slot. From all possible matching mechanisms encompassed under MDA\cite{Parsons}, we will choose the one presented on \cite{MDA}, which has already been used for energy allocation markets\cite{Alonso} and complies with certain interesting characteristics: 
\begin{itemize}
	\item Strategy-proof with respect to reservation price - Means that each agent's best strategy is to reveal truthful information regardless of what the other agents are doing, i.e.: no agent can benefit from lying about its price, which could hinder social welfare maximization.
	\item Weakly budget-balanced - All the payments between buyers and sellers sum to a positive value.
	\item Individually rational - The market encourages participation by ensuring non-negative profits.
	\item Asymptotically efficient - The market becomes more efficient with the increasing amount of participants, maximizing total profit obtained by all participants.
\end{itemize}

Economic transactions that impact the control of physical assets need to be managed in a secure, reliable and transparent manner. A centralized entity enforcing those requirements may imply transaction fees that could make energy micro-transactions unfeasible, due to infrastructure operation and maintenance costs; e.g.: for avoiding a single point of failure or Denial of Service attacks, to improve security and reliability, costly third party cloud computing resources may be needed. Furthermore, such entities generally require security deposits (collateral) to cope with counter-party risks, such as insolvency issues, that may lead to high costs of entry and to an increase in fees/taxes due to auditing costs.

The alternative to a centralized auctioneer would be implementing market matching, clearing and settlement rules into open standardized code to be executed by a network of distributed agents. Thus, we would still be under an institutional price-setting\cite{Parsons} mechanism but the institution could be implemented as a distributed application. Next we will analyse some of the challenges for building a distributed market involving digital and physical assets and we will present some of our implementation choices.

\subsubsection{Transaction Verification}
One of the main challenges of a digital market comes from the interactions with the physical world, particularly in terms of verification of transaction enforcement, which can hardly be done in a distributed fashion and generally requires a trusted entity. DSOs could be in charge of verifying the enforcement of transactions by measuring electricity injections and extraction flows, which is one of the main roles they currently play, they are trusted with and audited for. Furthermore, in the future, such verification could be done automatically as a service provided by smart meters.

As most transactions are for the exchange of resources on the future, they behave as options or forward contracts, and verification of service delivery is separated in time from transaction clearing\cite{Micromarkets}. As we cannot tag electrons to identify where they go, we cannot differentiate the injections and extractions that were transacted on the local market from the ones provided by the utility supplier. For instance, we will assume that the energy transacted on the local market has the preference/priority regarding verification. For example, if a household equipped with a PV sells its forecasted surplus of electricity in the local market, and then it produces less, the supplier will provide the energy to the consumer in the contract, but the producer will have to pay the difference in price. This way, agents are incentivized to have good forecasts and to transact accordingly.

\subsubsection{Blockchain-based transaction mechanism}
When two parties want to establish a digital transaction over a distributed network, they need to claim ownership of the digital/physical assets involved and to ensure transactions are non-reversible, particularly when digital transactions comprise non-reversible physical services. Digital signatures together with some asset registry, provide a solution to the problem of ownership and authorization for spending assets, but do not solve the double-spending problem. Double-spending is a fundamental problem of the cryptocurrency world, which occurrs when an entity that earns certain amount of digital assets transacts that same amount simultaneously to two other entities, and if there is no distributed way for determining which transaction arrived first, transactions can be easily reverted in the absence of a centralize clearing entity.

Bitcoin provided the first distributed solution to double spending problem by establishing a chain of ownership and a mechanism for agreeing in a distributed fashion about transfers of ownership and validity of transactions. This mechanism is called the blockchain, as it groups transactions into blocks and chains them together using cryptographic technologies, building up the transitions into current state of affairs. From those technologies, Proof of Work (PoW) is the most important, as it makes it difficult for a node to publish a new block of transactions by tying the validity of a new block to the solution of a complex mathematical puzzle. The solution to the puzzle requires trial and error and difficulty is adjusted for blocks to be published periodically. This makes it very hard and costly for an attacker to fork the chain and introduce invalid transactions, as it needs to have more than half of the hashing power of the network to carry an effective attack. As this mechanisms requires a lot of energy for maintaining the agreement on the network, there exist alternatives such as Proof of Stake (PoS), which are more efficient and may have similar security guarantees. Though, our focus is on the blockchain, because it can be seen as a distributed, self-authenticating, time-stamped store of data\cite{Eris}, which in addition allows to execute code in a distributed and extremely reliable manner, and thus can be used for many things other than cryptocurrencies.

Ethereum is the most relevant project generalizing the usage of blockchain technology by providing the necessary tools for the implementation of Smart contracts permanently stored on the blockchain. The Smart contract concept in the context of blockchain can be seen as a modular, repeatable script, which can be used to build distributed autonomous applications. When a transaction is sent to a contract the code is automatically executed and is able to use the data which is sent with the transaction\cite{Eris}. We plan to use the Ethereum blockchain as a base for the implementation and testing of our distributed energy balancing market.

\subsubsection{Blockchain-based distributed energy market}
We can translate the notion of double-spending, which means that a resource cannot be used twice, into the energy management case where the resource would be an energy block or certain flexibility budget, and cannot be owned by two entities at the same time, i.e.: at the neighborhood scale, energy transactions involving KWatts or NegaWatts should not be duplicated. In the energy case, we could build a chain of ownership on resources, that would be traded in advance in a per time-slot fashion, in exchange of some reward that would represent the utility of using such energy/flexibility, and that cannot either be double-spent.

We envisage the implementation of an exchange market for digital tokens which will represent the exchange of physical assets per digital currency. Money will be represented by a token we will refer to as Bitcoin, but could be any other digital currency. Physical assets, i.e.: energy blocks/flexibility budget, will be represented by a token we will call ecoin, whose economic value will be tethered to the verification of the corresponding electricity flow. There should be a distributed registry of ownership of ecoins for these to be available for transactions on the markets, which will enable to enforce security (authenticate participants) and privacy (assign pseudonyms). Trading agents must register availability of ecoins for each time slot, as the clearing procedure will verify the ecoins in order to validate transactions. Such registry can easily be implemented on the blockchain or on a distributed database with similar properties.

Initially, ecoins on the ownership registry will not have any economic value, but, using the fact that digital tokens are not completely fungible, each ecoin will acquire its value only once there is a proof of flow (PoF) over the contracted flow of energy involving those ecoins, meaning an ecoin is minted each time an energy exchange is physically verified\footnote{Similar to the SolarCoin Proof of solar production. \url{http://solarcoin.org/}}. Once there is the PoF, meaning once the DSO verified the enforcement of the contract and signed the transaction, the transaction is validated and the ecoins can be exchanged by Bitcoins at the rate agreed on the contract. The ecoins would represent a certificate of green energy consumption and can have many applications, such as social comparison/competition among electricity consumers for the energy efficiency or CO2 emission reductions.

There will be one or several smart contracts implementing each market, to which traders would need to address their bids and ask in order to participate. Those smart contracts would implement each aspect of the market, including matching offers (through an implementation of MDA), clearing (by verifying the buyer has enough Bitcoin and the seller owns enough ecoins), and settlement of transactions (registering on the blockchain the exchange of Bitcoins and ecoins at the rate specified by the matching mechanism).

In addition to the local balancing markets, the distributed nature of blockchain-based markets will enable any aggregator or supplier to establish a call for flexibility through the implementation of a smart contract, through which flexible DER owners can interact and transact their flexibility, without the need of having a previously signed legal contract. Furthermore, the transaction history could be used as a source of reputation of the different service providers for one side and on the different flexibility providers from the other side.

\paragraph{Benefits of Blockchain-based distributed markets}
\begin{itemize}
	\item A ledger of all transactions performed is publicly available, which favors transparency and accountability. 
	\item Requires minimal network infrastructure: Transactions can be sent over any network, including insecure ones, as transactions are signed and contain no confidential information\cite{Bitcoin}.
	\item It represents a generic infrastructure that can be seamlessly extended to other services (gas, water, etc.).
	\item The billing mechanism is embedded as part of the system, which reduces operation costs. 
	\item Flexibility to adapt to increasingly dynamic distribution grid transactions, with plug \& play characteristics.
	\item Reliability and availability - The ledger of all transactions performed is available on several nodes of the infrastructure which improves reliability as the infrastructure would seamlessly support node failures and communication errors.
	\item Copes with the issue of nodes on the network not necessarily being trustworthy.
\end{itemize}

\paragraph{Drawbacks / Challenges}
\begin{itemize}
	\item Proof of Work mechanism requires a lot of energy; which makes the distributed agreement mechanism inefficient. Thus, alternatives for proof of work need to be envisaged, such as Proof of Stake.
	\item Another issue is the amount of transactions per second that the network is capable of processing. (Bitcoin only supports 7 per second).
	\item Which is the minimum feasible value to be transferred on over the blockchain with respect to fees?  
	\item Other problem is the size of the blockchain, as it gets bigger as it is used, and less nodes are capable of managing the bloated blockchain. 
\end{itemize}

\subsection{Household control level}
The implementation of a HEMS imposes less stringent requirements in terms of communications and computation than a neighborhood EMS, due to fewer devices and smaller size of the problem. Thus, we will assume the intelligence for household energy management will be centralized in a dedicated controller. In a similar way to energy management, there will be other dedicated controllers for providing services on different domains such as health, security, entertainment, etc. Those controllers influence the environment through control transactions sent to actuators. Such processes pursue different objectives but may share the same set of actuators or the same environmental scope through different actuators without further coordination. The lack of orchestration among controllers can create conflicts that will undermine interoperability and increase complexity of Smart Home Services.

Issues with conflicting digital transactions are generally solved by putting trust in a single entity in charge of centralizing control, keeping a global ledger of the state of devices and clearing transactions for avoiding conflicts. An example of such an entity is \textit{Works With Nest}\footnote{https://nest.com/works-with-nest/} platform, which maintains a centralized document with the state of all resources involved. The document is synchronized among partner entities and allows to implement simple ECA-like rules for controlling the behavior of connected devices. Simultaneously, \textit{Works With Nest} enable households to participate to rebates proposed by utilities in exchange for reducing electricity consumption. The platform orchestrates thermostats to avoid conflicts between pre-determined rules and the enforcement of commitments with utilities.

A centralized orchestrator managing every conflicting situation favours siloed ecosystems, each of which will implement its own architecture undermining interoperability and hindering value creation. Such an approach lacks plug \& play capability, which is fundamental for an heterogeneous environment over lossy networking infrastructure. Then, we need a distributed mechanism that would enable solving conflicting access to resources applicable to any domain, i.e.: a generic mechanism. A distributed agreement mechanism based on blockchain technology and self-enforcing smart contracts is envisaged to natively enforce security of transactions and to address the orchestration issues without relying on a central entity. Such a distributed agreement mechanism would allow to decouple controllers from end devices, through an horizontal open infrastructure where devices from all product manufacturers can participate.

\subsubsection{A chain of ownership to avoid conflicts}
We can model the conflicts problem as a double-actionning of a resource, where the resource is the control over a device, a set of devices or an environmental variable (e.g.: temperature) under certain context, i.e.: during a certain period of time. In order to solve the conflict, a contract is established to avoid two entities sharing the control of a certain resource at the same time.  The contract structures the acceptable means of exchange of ownership over resources; conditions for transferring ownership are previously stated (e.g.: at device deployment time \cite{Conflict}) and automatically enforced. Once a controller sends a transaction acquiring the control of a device, it can be sure of being in power of the resource as any transaction to control the device by another controller will be considered invalid by all nodes.

\subsubsection{On-premises ECA distributed implementation}
In addition to solving conflicts, such a distributed system based on state of the art cryptographic primitives can provide further benefits: the blockchain-based platform can enable enforcement of ECA rules entirely on-premises. Most of the manual and ECA control would be handled locally and autonomously by smart contracts running over the ``unmanaged'' blockchain. This will improve security and privacy with respect to cloud-based ECA implementations. This will also enable connected device vendors to avoid paying, or to reduce the bill (reduced usage), for cloud infrastructure along the lifetime of the device (expected to last several years).

\subsubsection{Challenges of a household blockchain}
Given that a cryptocurrency makes no sense in the interactions among smart home devices, one of the biggest challenges is to ensure the effectiveness of the consensus algorithm and the participation of devices on the network, which can be undermined by lack of incentives (for device manufacturers to make their devices blockchain-enabled). In the household we do not need censorship resistance virtual cash or proof of work based systems, but we do have a network-based sybil problem. We are analyzing the use a Proof of Stake mechanism and to establish a reputation token instead of a digital currency. We envisage the reputation coin as an utilitarian mechanism to assess the value that device manufacturers offer to users by actively participating in securing the blockchain network, and thus enabling interoperability and distributed agreement.

Other challenges that would need to be analyzed by further work are the following:
\begin{itemize}
	\item The delay imposed by the network and the difficulty to publish a new block (time between blocks) are determinant to analyze the feasibility of using distributed agreement mechanisms for real-time applications at a household level, i.e.: the possibility of enabling micro-transactions at a speed fast enough to provide ECA rule-based services, involving user interaction.
	\item Blockchain-based distributed agreement are not continuously consistent as there will often exist forks from the main blockchain which will require some time window for the "voting" to happen and to be back in a "synchronous" state. The impacts of such forks on the interaction among devices and controllers needs to be assessed and minimized.
\end{itemize}

\section{Conclusions}
Major ongoing evolutions in the electricity industry potentially represent a key driver for the ``energy transition'' and related objectives. Nevertheless, they impose new challenges in the whole value chain. The required transformations of the Distribution grid are today delaying, and in some cases blocking, the possibility to leverage key opportunities. In this paper we propose a new paradigm, that we named Virtualized Distribution Grids, which facilitates implementing the required solutions for major distribution grid challenges without requiring important infrastructure investments. The proposal includes a new approach for designing DSOs' Energy Management Systems. The hierarchical architecture we present enables the coordinated participation of any type of player, including DSOs, aggregators, and end users (that become \textit{prosumers}). In addition, in this general framework, we propose specific \textit{market-based} solutions that enable deploying advanced technologies (local production, storage, BEMs, IoT based demand response systems and other on-premises technologies), by different players, and coordinating those players in a way to optimize the overall value while keeping the distribution network stable and providing the expected
quality of supply. In addition, we present a distributed architecture, based on the blockchain principles, that supports the implementation of the proposed markets. Finally, we extend the architecture to solve key challenges raised in smart homes, beyond energy management, including policy-based coordination of controllers from independent service providers acting on the same connected devices.

We are working in the development and deployment of the proposed solution over a test bed composed of real residences and on the evaluation of various market approaches, starting with real-time balancing markets and DSO reserve mechanisms.

\bibliographystyle{hieeetr}
\bibliography{references}

\end{document}